\documentclass[10pt,twocolumn,letterpaper]{article}

\usepackage{iccv}              

\usepackage{upgreek}

\usepackage{pifont}
\usepackage{times}
\usepackage{microtype}
\usepackage{epsfig}
\usepackage{caption}
\usepackage{longtable} 
\usepackage{float}
\usepackage{placeins}
\usepackage{color, colortbl}
\usepackage{stfloats}
\usepackage{enumitem}
\usepackage{xstring}
\usepackage{multirow, makecell}
\usepackage{xspace}
\usepackage{url}
\usepackage{subcaption}
\usepackage{xcolor}
\usepackage{algorithm}
\usepackage{algpseudocode}
\usepackage[hang,flushmargin]{footmisc}
\usepackage{tabularx}
\usepackage{threeparttable}
\usepackage{amsmath}
\usepackage{amssymb}
\usepackage{xspace}

\makeatletter
\algdef{SE}[FOR]{ForEach}{EndForEach}[1]{%
  \algorithmicfor\ #1\ \algorithmicdo
}{%
  \algorithmicend\ \algorithmicfor
}
\makeatother

\usepackage{graphicx}
\usepackage{bm}
\usepackage{booktabs}
\usepackage{bigstrut}
\usepackage{bbding}
\usepackage[misc]{ifsym}

\definecolor{iccvblue}{rgb}{0.21,0.49,0.74}
\usepackage[pagebackref,breaklinks,colorlinks,allcolors=iccvblue]{hyperref}

\def\paperID{} 
\def\confName{ICCV}
\def\confYear{2025}

\title{Learning to See Through Flare}

 \author{
    Xiaopeng Peng\textsuperscript{1}, 
    Heath Gemar\textsuperscript{2}, 
    Erin Fleet\textsuperscript{2},
    Kyle Novak\textsuperscript{2},
    Abbie Watnik\textsuperscript{2}, 
    Grover Swartzlander\textsuperscript{1}\\
  \textsuperscript{1}Rochester Institute of Technology, 
  \textsuperscript{2}U.S. Naval Research Laboratory
}

\begin{document}
\maketitle


\begin{abstract}
Machine vision systems are susceptible to laser flare, where unwanted intense laser illumination blinds and distorts its perception of the environment through oversaturation or permanent damage to sensor pixels. We introduce NeuSee, the first computational imaging framework for high-fidelity sensor protection across the full visible spectrum. It jointly learns a neural representation of a diffractive optical element (DOE) and a frequency-space Mamba-GAN network for image restoration. NeuSee system is adversarially trained end-to-end on 100K unique images to suppress the peak laser irradiance as high as $10^6$ times the sensor saturation threshold $I_{\textrm{sat}}$, the point at which camera sensors may experience damage without the DOE. Our system leverages heterogeneous data and model parallelism for distributed computing, integrating hyperspectral information and multiple neural networks for realistic simulation and image restoration. NeuSee takes into account open-world scenes with dynamically varying laser wavelengths, intensities, and positions, as well as lens flare effects, unknown ambient lighting conditions, and sensor noises. It outperforms other learned DOEs, achieving full-spectrum imaging and laser suppression for the first time, with a 10.1\% improvement in restored image quality.
\end{abstract}
\section{Introduction}
\label{sec:intro}
\begin{figure}[t] 
	\begin{center} 
	\includegraphics[width=1.0\linewidth]{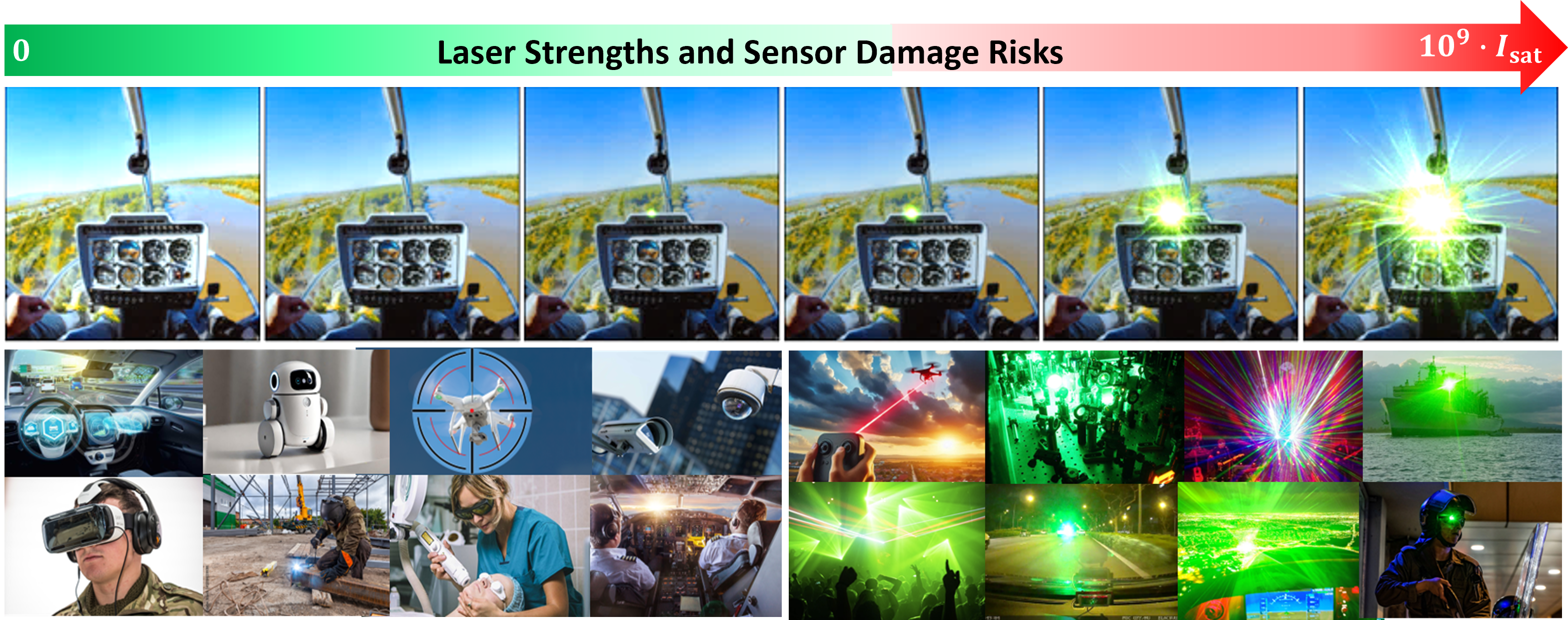} 
	\end{center} 
	\caption{Illustration of sensor damage risks under the laser illumination (row 1) and applications of sensor protection (row 2), which includes mitigating disruption and improving vision performance of autonomous vehicles, robots, consumer cameras, as well as mixed reality devices for eye protection in scientific experiments, aviation, medical treatment, entertainment laser shows, and more.} 
	\label{fig:Intro}
\end{figure}

Continuous advancements of laser technology have enabled the ready availability of low-cost, compact, and powerful lasers which, if misdirected toward an image sensor, may cause objectionable dazzle (e.g., sensor saturation and lens flare) or irreversible anomalies. As illustrated in Fig.\ref{fig:Intro}, lasers can disrupt vision and mislead the tracking system of unmanned aerial vehicles \cite{steinvall2021potentialp2,steinvall2023laser,lewis2023disruptive}. Adversarial laser attacks against the sensor of autonomous or robotic vehicles have been demonstrated to significantly compromise their safety and reliability \cite{duan2021adversarial,kim2022engineering, sun2024embodied}. Lasers also present risks to sensors in mixed reality devices (e.g., video see-through head-mounted displays). These devices may advance the development of eye protection goggles \cite{quercioli2017beyond, owczarek2021virtual, deniel2022occupational, li2023mixed} by providing high-quality imaging capability that may otherwise not be possible. They are crucial in scientific experiments, aviation \cite{FAA2023laserincident}, law enforcement, manufacturing \cite{OSHA2023laserhazard}, medical treatment \cite{malayanur2022laser}, and more. In addition, laser-induced damage to consumer camera sensors has been reported during entertainment events, such as laser shows \cite{ILDA2022lasershow}.

The laser-induced saturation and damage of an imaging sensor depend on both the sensor and the laser characteristics. The damage thresholds of silicon-based imaging sensors are typically six to nine orders of magnitude higher than their saturation thresholds \cite{ritt2019laser,theberge2022damage}. To mitigate laser-related sensor risks and image degradation, optical techniques such as wavelength multiplexing \cite{ritt2019preventing, ritt2020use}, holographic coatings \cite{caillieaudeaux2024thermoset}, liquid crystals \cite{wang2016self,zhang2023advanced}, metamaterials \cite{howes2020optical, bonod2023linear}, integration time reduction \cite{lewis2019mitigation}, smoke obscurants \cite{schleijpen2021smoke}, and analytical phase masks \cite{ruane2015reducing, luo2024performance}. Learned diffractive optical elements (DOEs) have also been explored, but their performance is typically constrained by limited field-of-view or spectral bandwidth \cite{wirth2020half, peng2022computational, peng2024learning, zhang2024learned, luo2025opsecurecam, meyer2025laser}. None of these approaches has been found to simultaneously satisfy the desired bandwidth, response time, dynamic range, stability, and image quality.


In this work, we present NeuSee, a computational imaging framework that jointly learns a neural representation of a diffractive optical element (DOE) and a Mamba-GAN-based image restoration network operating in the frequency domain. NeuSee is trained end-to-end in an adversarial manner on 100K unique images, enabling it to suppress peak laser irradiance up to $10^6$ times the sensor saturation threshold $I_{\textrm{sat}}$, a level that could otherwise cause sensor damage without the use of a DOE. The framework comprises a learnable neural phase mask that attenuates laser light while preserving the transmission of the background scene. It also features a novel restoration network that corrects multiple image degradations, such as inpainting the saturated area, reducing image blur, and removing noise, resulting in a high-quality image. The contributions are summarized as follows.

\begin{itemize}
\item We present the first framework that jointly learns a diffractive optical element (DOE) and image restoration to protect sensors from high-energy lasers across the full visible spectrum.  

\item The learned DOE achieves both high background-light throughput and laser suppression. A neural DOE representation with state-space GAN restoration and a two-stage learning strategy disentangles conflicting objectives.  

\item We curate a 100K high-resolution dataset with a physics-based synthesis pipeline, generating diverse training images across laser spectra, intensities, positions, ambient illumination levels, and noise.  

\item Our system outperforms existing sensor-protection DOEs in laser suppression ratio, imaging spectrum, laser spectral coverage, and image quality.  



\end{itemize}

\section{Related Work}
\label{sec:related-work}
\emph{End-to-End Learned Camera System}. Computational imaging and photography are emerging areas that focus on improving and extending the capabilities of traditional imaging and camera systems using optical and computational methods. By altering light transmission at the pupil plane using an amplitude mask or a phase mask, the coded aperture approach has been investigated in many applications, such as coded exposure \cite{raskar2006coded}, achromatic imaging\cite{peng2016diffractive, dun2020learned},  high dynamic range \cite{sun2020learning, metzler2020deep}, lens glare suppression \cite{raskar2008glare, rouf2011glare}, extended depth of field \cite{sitzmann2018end, wu2019phasecam3d, tan2021codedstereo}, light field imaging \cite{veeraraghavan2007dappled, peng2016shape}, lensless imaging \cite{boominathan2022recent}, privacy-preserving imaging \cite{hinojosa2021learning, tasneem2022learning}, granular imaging \cite{peng2016randomized, peng2014mirror}, hyperspectral imaging \cite{jeon2019compact,baek2021single}, AR glasses \cite{mou2024p, mou202438, yang2025rf} and more. Instead of relying solely on separately learned optics, recent advances in computational imaging have embraced end-to-end optics learning \cite{sun2020end, shi2022seeing, tseng2021differentiable, li2023extended}, where phase masks are co-optimized with image restoration algorithms. This approach integrates differentiable simulations of optical systems with task-driven neural reconstruction models, enabling greater adaptability and system-level optimization. Unlike traditional designs that primarily target optical aberrations in isolation, end-to-end methods consider the full imaging pipeline, tailoring both optics and algorithms to the specific task. The resulting phase masks learn to encode latent image information into PSFs, which is subsequently decoded through the reconstruction network.  



\begin{figure*}[tb] 
	\begin{center} 
	\includegraphics[width=0.98\linewidth]{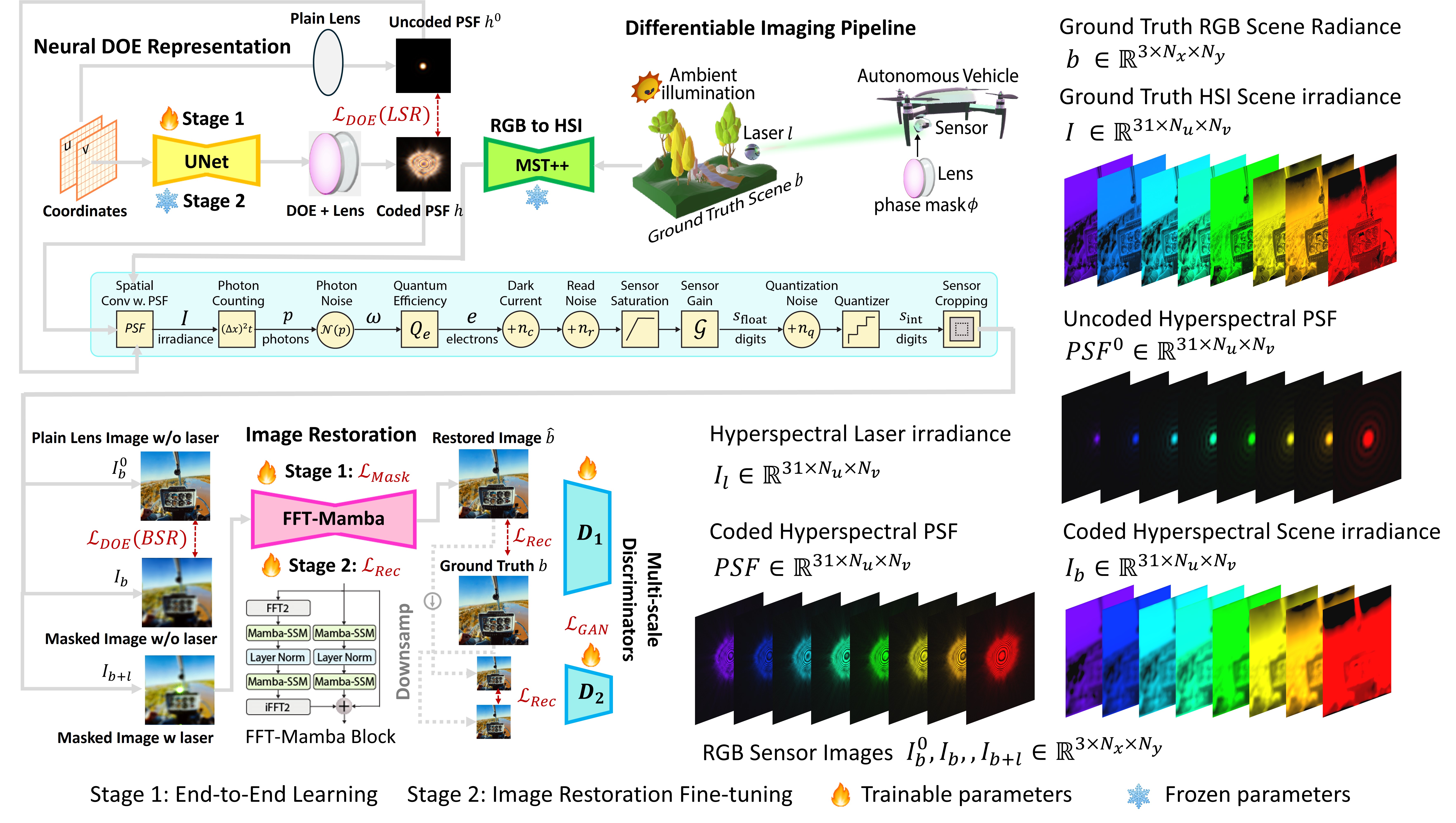} 
	\end{center} 
    \vspace{-0.5cm}
	\caption{The proposed NeuSee system is a jointly learned imaging framework designed to protect sensors from laser dazzle across the full visible spectrum. Our system involves \textbf{1) A physics-based} differentiable imaging pipeline. To achieve accurate simulation of diverse scene and lasers, pre-trained hyperspectral (HSI) reconstruction net is used to transform each RGB radiance map to a 31-band HSI volume; \textbf{2) UNet-based neural representation} of the DOE (or phase mask); \textbf{3) A novel Mamba-GAN  architecture} for image restoration, which makes use of image intensity and frequency information. The training of NeuSee includes two stages. \textbf{Stage 1} trains the DOE jointly with Mamba-GAN restoration with DOE loss and GAN loss. In \textbf{Stage 2}, the quality of the restored image is finetuned by frozen the DOE and train the Mamba-GAN with reconstruction and GAN loss.} 
	\label{fig:pipeline}
\end{figure*}

\emph{Image Restoration}. Learning a direct mapping from the sensor image $s$ to the background scene radiance $\hat{b}$ using DNN has been extensively studied for many low-level computer vision tasks \cite{zhang2022deep}, including image deblurring, denoising, deraining, dehazing, in- and outpainting, as low light, and many computational imaging applications \cite{peng2021cnn, peng2022computational, peng2017randomized, peng2015image}. Vision transformers \cite{liang2021swinir,wang2022uformer,zamir2022restormer} seek to capture long-range image dependencies using self-attention mechanisms. Alternatively, the spatial gating unit \cite{liu2021pay} was introduced with the MLP to achieve a performance comparable to that of transformers \cite{tu2022maxim}. Recent advances in state-space sequence models (SSMs or Mamba) \cite{gu2023mamba} also demonstrated strong efficiency and effectiveness  in vision tasks \cite{zhuvision, ma2024u,zhou2025neural}. Additionally, changing the image basis through linear or nonlinear transforms to better suit the orthogonally preconditioned optimizers was found to accelerate the convergence and improve the performance of the DNNs. One notable finding involves the use of FFT features to correct spectral bias and improve the learning of high-frequency functions from low dimension data \cite{rippel2015spectral, chi2020fast, tancik2020fourier}.
Furthermore, by breaking down a complicated task into subproblems and solving them progressively, multistage and multiscale frameworks allow supervision and feature fusion in multiple restoration stages and multiple image scales\cite{zhang2019deep, zamir2021multi} and encourage the recovery of image details. Embedding of kernel functions \cite{feng2021removing} and image coordinates \cite{liu2018intriguing, lin2019coco, lin2021infinitygan} into neural networks has also been introduced, respectively, to utilize the knowledge of system PSF and positional information of an image.  

Generative models seek to learn the joint distribution $\mathrm{Pr}(s,b)$, which allows for more accurate recovery of missing data such as image saturation. Unlike VAEs \cite{kingma2014auto} and flows \cite{kingma2018glow}, GANs do not rely on an explicit inference model. Instead, they learn  the target distribution from an input distribution by seeking a Nash equilibrium between a generator and a discriminator through a minimax game. Diffusion models \cite{sohl2015deep, rombach2022high} learn the implicit latent structure of a dataset by modeling the way in which data points diffuse through the latent space. Generative methods face the performance trillema \cite{xiao2021tackling} of sampling quality, diversity, and speed. GAN models outperform many VAEs and flows in generating realistic looking images. While diffusion models can generate high-quality images with improved sample diversity \cite{dhariwal2021diffusion, ho2021classifier}, their potential in real-time applications is limited due to the high cost of iterative sampling. The single-step distilled diffusion models remain underperforming in sample quality compared to GANs \cite{zheng2023fast, meng2023distillation}.

\section{Image Formation Model}
Physics-based modeling enables accurate characterization of imaging systems and supports end-to-end optimization of optical elements and restoration algorithms, with efficient transfer from simulation to real data. We next outline an image formation model based on wave propagation.  

\subsection{RGB to HSI Irradiance}
While proven effective in other computational imaging tasks \cite{shi2022seeing, tseng2021differentiable}, ensuring continuous laser coverage is critical for our application. Consequently, simulating point spread functions (PSFs) and sensor responses using only three RGB channels is insufficient. To overcome this limitation, and given the scarcity of hyperspectral imaging (HSI) data, we employ a pre-trained spectral reconstruction network, $G_{hsi}$ (MST++ Net \cite{cai2022mst++}), to recover 31-channel scene irradiance $b(x,y,\lambda)$ in the visible spectrum (400–700 nm) from RGB ground-truth irradiance $b(x,y,3)$. This approach enables us to leverage the broader availability of RGB ground-truth datasets. The transformation is formulated as:

\begin{equation}
b_{\lambda}(x,y,31) = G_{hsi}\big(b(x,y,3)\big)
\label{eq:HSI}
\end{equation}


\subsection{Neural Representation of the DOE}
In this study, the incident laser is characterized as a plane wavefront at the entrance pupil.  To achieve sensor protection, we introduce a learnable diffractive optical element (DOE)  
at the entrance pupil of the system, the height of which is written as a mapping of the pupil Cartesian coordinates of pupil plane $(u,v)$ by a phase representation net $G_{rep}$:
\begin{equation}
h_{DOE}(u,v) = G_{rep}(\mathrm{concat}(u,v))
\end{equation}

\noindent here we use an 8-layer UNet as the phase representation net. Assuming the DOE has a wavelength dependent refractive index $\Delta n(\lambda)$, the phase distribution of the DOE is given by:
\begin{equation}
\phi_{DOE}(u,v) = 
\frac{2\pi}{\lambda}\cdot\Delta n(\lambda)\cdot h_{DOE}(u,v) 
    \label{eq:phi_doe}
\end{equation}
 
\subsection{System Point Spread Function}
\label{ssec:imaging-formation-psf}
The spectral irradiance distribution in the focal plane $(x,y)$ is described by wavelength-dependent pupil function and point spread function (PSF):

\begin{subequations} \label{eq:irrad_scene} 
\begin{align} 
& U_{\lambda}(x,y,\lambda) = \frac{e^{ikf}}{i\lambda f} \iint A(u,v)e^{i\phi(u,v)}e^{ik(xu+yv)/f}\mathrm{d}u\mathrm{d}v \\
& PSF_\lambda(x,y, \lambda) = \bigg|U(x,y,\lambda) \bigg|^2
\end{align}  
\label{eq:PSF_coded}
\end{subequations}

\noindent where $f$ 
is the focal length 
of the imaging lens,
$k=2\pi/\lambda$ is the wave number,
and the integration is made over the interior of a circular
aperture $A(u,v)$ of radius $R$.

\begin{figure}[t] 
	\begin{center} 
	\includegraphics[trim=0cm 0.8cm 0 0, width=1.0\linewidth]{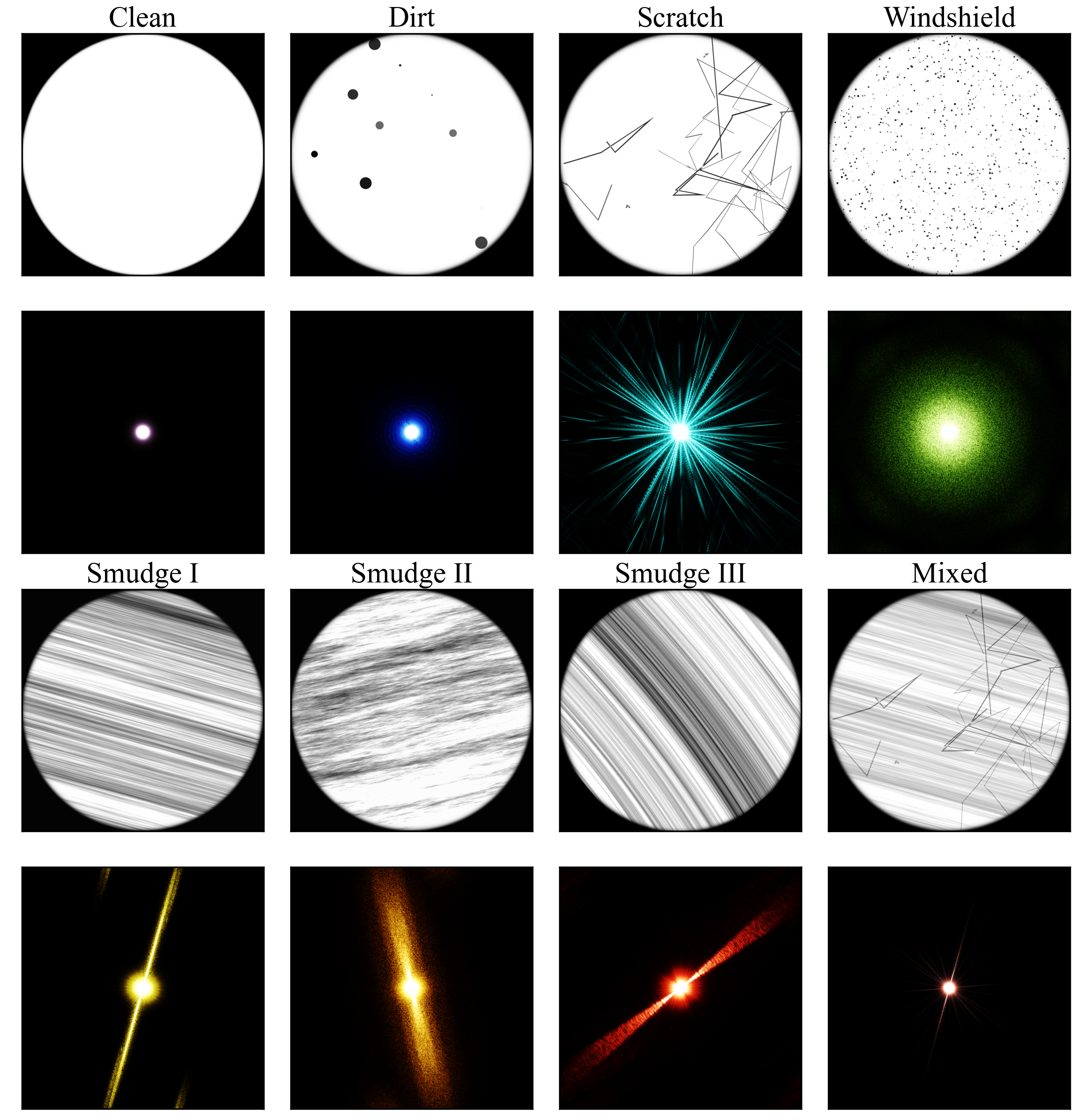} 
	\end{center} 
	\caption{Lens flare effects under laser illumination at various wavelengths, caused by lens imperfections such as dirt, scratches, windshield dents, grease, smudges, or their combinations.} 
	\label{fig:LensFlare}
    \vspace{-0.5cm}
\end{figure}

To accelerate the computation of Eq. \ref{eq:PSF_coded}, we make use of the efficient Scaled Fresnel method \cite{scalefft2022} that relates the Fourier transform of the field in the pupil with that at the sensor planes when the two domains have different pitch of the pixels. When a camera captures a scene with a strong light source in presence, the resulting image may exhibit lens flare artifacts that caused by light scattering. These flares can manifest in various forms, such as halos, streaks, color bleeding, and haze. They are typically caused by dirt, scratches, windshield dents, grease, or smudges, or a mixture of them. Inspired by previous work \cite{hullin2011physically, wu2021train, dai2022flare7k}, we also simulate these lens flare effects as shown in Fig. \ref{fig:LensFlare}.

\begin{table}[tb]
\centering
\captionsetup{justification=centering}
\caption{Physical Parameters}
\begin{tabular}{lcc}
\toprule
\makecell{Parameter} & \makecell{Symbol} & \makecell{Value} \\
\midrule
Background wavelength    & $\lambda_b$              & 400-700 nm  \\
Laser central wavelength & $\lambda_{l}$          & 400-700 nm  \\
Effective focal length   & $f$                      & 0.11 m \\
Exposure time            & $t$                      & 0.1 sec\\
Aperture diameter        & $W_a$                    & 11 mm \\
Quantum efficiency       & $Q_e$                    & 0.56\\
Sensor gain              & $\mathcal{G}$            & $0.37$\\
Full well capacity       & $e_\mathrm{sat}$         & $25500e^-$\\
Read noise (mean)        & $\mu_{r}$                & $390e^-$\\
Read noise (std.)        & $\sigma_{r}$             & $10.5e^-$\\
Dark current             & $\sigma_{c}$             & $0.002e^-$\\
Bit depth per channel    & $\mathrm{bpc}$           & $16$\\
Pupil pitch              & $\Delta u = \Delta v$    & 3.74 $\upmu$m\\
Sensor pitch             & $\Delta x = \Delta y$    & 2.9 $\upmu$m\\
Pupil resolution         & $N_u \times N_v$         & $2160 \times 2160$\\
Sensor resolution        & $N_x \times N_y$         & $2048 \times 2048$ \\
\bottomrule
\end{tabular}
\vspace{-0.2cm}
\label{tab: params_phys}
\end{table}

\subsection{Sensor Image}
\label{ssec:imaging-formation-image}

\begin{figure}[t] 
	\begin{center} 
	\includegraphics[trim=0cm 0.8cm 0 0, width=1.0\linewidth]{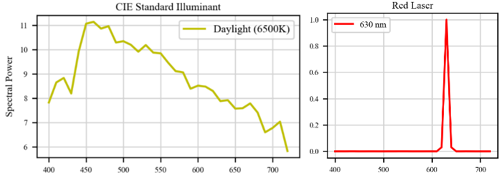} 
	\end{center} 
	\caption{Spectral profile of scene illumination and a red laser.} 
	\label{fig:spectral_curve}
    \vspace{-0.2cm}
\end{figure}

\emph{Irradiance Distribution}. A shift-invariant imaging system integrates the radiance distribution over the solid angle that is extended by the aperture through spatial convolution $*$, resulting in an irradiance map in the image plane. It is assumed that the background illumination has a visible wavelength $\lambda_b$ and the laser has a narrow wavelength $\lambda_l$. For the phase-coded system, we represent the irradiance distributions in the sensor plane of the background scene and a laser spot:
\begin{subequations} \label{eq:irrad_scene} 
\begin{align} 
&I_{b}(x,y,\lambda) = \mathcal{F}^{-1}(\mathcal{F}(b_\lambda) \cdot U_\lambda)\cdot T_b(\lambda)\\
&I_{l}(x,y,\lambda) = \delta(x-\Delta l_x, \; y-\Delta l_y) \cdot PSF(x,y,\lambda) \cdot T_l(\lambda)
\end{align}  
\end{subequations}

\noindent where 
$b_\lambda(x,y)$ is the ground truth radiance map of the background scene,
$T_b(\lambda)$ is the simulated CIE daylight spectral curve 
shown in Fig. \ref{fig:spectral_curve}, $T_l(\lambda)$ represents the 
spectral profile of the laser characterized by a Gaussian distribution centered at the wavelength $\lambda_l$ and having a full width at half maximum bandwidth $\Delta \lambda_\text{FWHM} =$ 10 nm.The laser is considered a Dirac Delta function, which targets the sensor at a normal $\vec{n}_l=(n_{u}, n_{v})$ with respect to the optical axis. Its footprint shift on the focal plane is thus given by $(\Delta l_x, \Delta l_y) = f\cdot \vec{n}_l$. $\mathcal{F}$ and $\mathcal{F}^{-1}$ represents respectively the Fourier and its inverse transform, and $\delta(\cdot)$ represents the Dirac delta function, where we assume the values of $b_\lambda(x,y)$ are normalized to the range [0,1]. By replacing the coded PSF $h$ with the uncoded PSF $h_{0}$ in Eq.\ref{eq:irrad_scene}, the irradiance maps of the background scene and the laser are defined as $I_{b0}$ and $I_{l0}$ respectively for an unprotected system. 









\bigskip

\emph{Sensor Saturation.} For a given wavelength, the irradiance value that saturates a sensor is expressed as: 
\begin{equation}
I_\mathrm{sat} (\lambda)= e_\mathrm{sat}\cdot \frac{\mathfrak{h}\cdot c}{\lambda\cdot t \cdot(\Delta x)^2 \cdot Q_{e} }
\label{eq:ThresSat}
\end{equation}
\noindent where $e_\mathrm{sat}$, $Q_{e}$, and $\Delta x$ are respectively the full well capacity, quantum efficiency, and pixel pitch of the sensor, $\mathfrak{h} = 6.63\cdot 10^{-34}$ $[\text{J} \cdot \text{s}]$ is Plank's constant,  $c = 3\cdot 10^8$ [m/s] is the speed of light in vacuum, and $t$ is the exposure time. For an unprotected system, let us express the peak irradiances of the background scene and the laser spot, respectively, as $I_{b0,\mathrm{peak}}$ and $I_{l0, \mathrm{peak}}$, which are proportionate to the irradiance saturation value:
\begin{subequations} \label{eq:peak_irrad_coded} 
\begin{align} 
& I_{b, \mathrm{peak}} = \alpha_b \cdot BSR \cdot I_\mathrm{sat}(\lambda_b)
    \label{eq:peak_irrad_background_coded} \\
& I_{l, \mathrm{peak}}=\alpha_{l} \cdot LSR \cdot I_\mathrm{sat}(\lambda_l)
    \label{eq:peak_irrad_laser_coded}
\end{align}  
\end{subequations}
\noindent
where $\alpha_{b}$ and and $\alpha_l$ are respectively the strength of the background illumination and the laser. The laser saturates a single pixel when $\alpha_l = 1$ and multiple pixels when $\alpha_l > 1$.  
Sensor damage may occur at $\alpha_l > 10^6$. For an optical system protected by a pupil plane phase mask the corresponding peak irradiance values of the background scene
and laser are respectively scaled by a background suppression value $BSR$ and 
the laser suppression value $LSR$. The values $BSR \sim 1$ and $LSR << 1$ lower the risk of sensor saturation and damage while maintaining the transmission rate of the scene irradiance. 

\emph{Photon to Electron.} Photons arrival at a sensor has a Poisson distribution, the rate of which is determined by the image irradiance $I$, the pixel pitch $\Delta x$, wavelength $\lambda$, and the integration time $t$:
\begin{equation}
p (\lambda) = \frac{(I_b + I_l)\cdot\lambda\cdot t\cdot(\Delta x)^2}{\mathfrak{h}\cdot c}
\label{eq:photon_count}
\end{equation}

\noindent The hyperspectral photons are then converted to RGB space via the following equation
\begin{equation}
p = \begin{bmatrix}
p_{R} \\[6pt]
p_{G}\\[6pt]
p_{B}
\end{bmatrix}
= k \sum_{i=1}^{N} S(\lambda_i)\, I(\lambda_i)\,
\begin{bmatrix}
\overline{x}(\lambda_i) \\[6pt]
\overline{y}(\lambda_i) \\[6pt]
\overline{z}(\lambda_i)
\end{bmatrix}
\, \Delta\lambda
\end{equation}

\noindent
where $N = 31$ 
\[
\begin{cases}
S(\lambda_i) : \text{spectral reflectance or radiance at wavelength } \lambda_i, \\[6pt]
I(\lambda_i) : \text{spectral power distribution (illuminant) at } \lambda_i, \\[6pt]
\overline{x}(\lambda_i),\, \overline{y}(\lambda_i),\, \overline{z}(\lambda_i) : \text{CIE 1931 color matching functions}, \\[6pt]
k : \text{normalization constant, } 
k = \dfrac{100}{\sum_{i=1}^{N} I(\lambda_i)\, \overline{y}(\lambda_i)\, \Delta\lambda}, \\[10pt]
\Delta\lambda : \text{wavelength sampling interval}.
\end{cases}
\]

\noindent  According to the central limit theorem, the Poisson distribution may be approximated by a Gaussian distribution, which was found to be a better characterization of our sensor in practice. The Gaussian distributed photon is given by $\omega\sim \mathcal{N}(c_1\cdot \mu_{p},  c_2\cdot \sigma_{p} )$, where its mean and standard deviation are written respectively as the modulated mean ($\mu_p$) and standard deviation ($\sigma_p$) of the photon arrival rate $p$, and $c_1$ and $c_2$ are the modulation coefficients.  Given a quantum efficiency $Q_e$, the collected  photons are converted to electrons: $e = Q_e \cdot \omega$,  followed by noise corruptions and the digitization process:
\begin{equation}
s=\mathrm{crop}\Bigg(\min\bigg(s_{\mathrm{sat}},\Big\lfloor\mathcal{G}\cdot\min\big(e_\mathrm{sat}, e + n_d + n_c\big)+ n_q\Big\rfloor\bigg)\Bigg)
\label{eq:SensorImage}
\end{equation}

\noindent where unwanted electrons generated by other factors are modeled as additive dark current $n_c$ and read noise $n_r$. The dark current has a Poisson distribution $n_c\sim \mathcal{P}(\mu_c)$, and the read noise is Gaussian distributed $n_r \sim \mathcal{N}(\mu_r, \sigma_{r})$. The mean values $\mu_c, \mu_r$ and the standard deviation $\sigma_r$ of the noise are obtained through sensor calibration.

\emph{Digitization.} Electrons are converted into an array of integer digital counts that represents the image recorded by the sensor. The total number of electrons that exceeds the full well capacity of the sensor $e_\mathrm{sat}$ is clipped. Electrons are then amplified by a sensor gain $\mathcal{G}$, producing an array of floating points. Uniformly distributed quantization noise $n_q\sim \mathcal{U}(-0.5, 0.5)$ is added to these digits, and floating-point digital values are then quantized to integer digital counts. The upper limit of the digital counts is determined by the bit depth per channel (bpc) of the sensor, where $\mathrm{s_{sat}} = 2^{\mathrm{bpc}}-1$. The size of the image recorded by the sensor is determined by the finite size $(W_s, H_s)$ of that sensor. Consider that a radiance map in the object plane has a size $(W_o, H_o)$ and the system PSF has a width $W_{h}$, the size of the image formed in the focal plane is given by $(W_o+W_{h}, H_o + W_{h})$. Boundary areas that exceed the sensor size are cropped. The values of the physical parameters used in the simulation match the experiment (see Table \ref{tab: params_phys}).


\begin{figure*}[t] 
	\begin{center} 
	\includegraphics[width=1.0\linewidth]{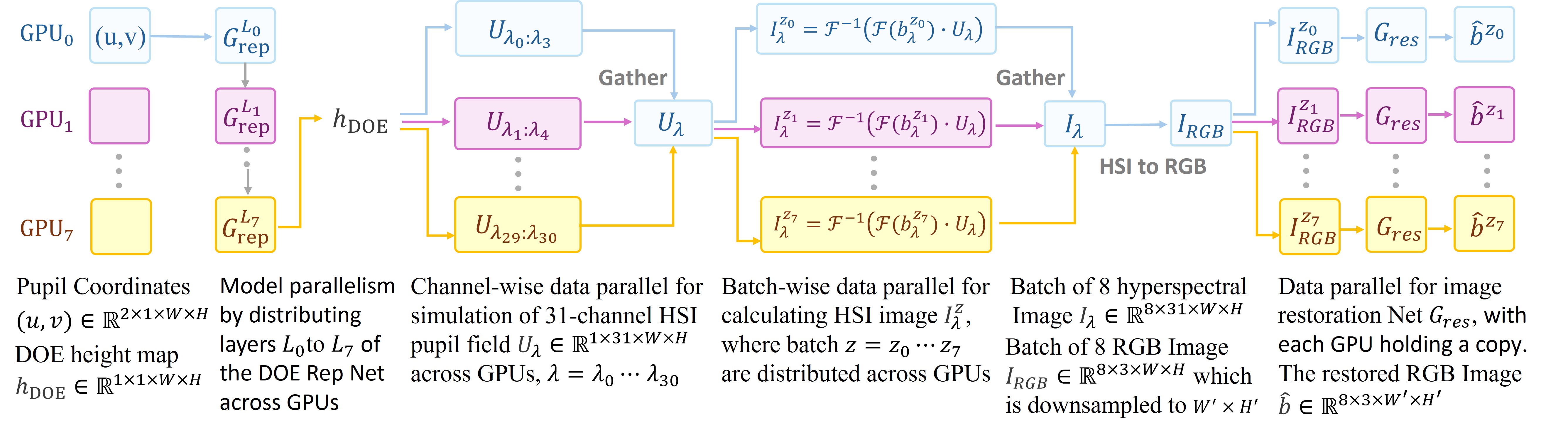} 
	\end{center} 
    \vspace{-0.7cm}
	\caption{Illustration of heterogeneous data and model parallelism for distributed end-to-end training of the NeuSee system.} 
	\label{fig:para}
    \vspace{-0.3cm}
\end{figure*}

\section{Image Restoration}
Here we introduce a Mamba-GAN for image restoration in frequency space, where the background scene radiance $\hat{b}$ is restored from the sensor image $s$:
\begin{equation}
\hat{b} = \mathrm{G}_{res}\big(s\big)
\label{eq:image_restore_model}
\end{equation}
\noindent The image restoration generator $\mathrm{G}_{res}$ and phase representation net $\mathrm{G}_{rep}$ are adversarially trained with a multiscale discriminator $D= \{D^{L}| L = 0,1\}$. As shown in Fig. \ref{fig:pipeline}, the architecture of $\mathrm{G}_{res}$ consists of 8 layers FFT-Mamba blocks.

To better recover image details, a coarse-to-fine architecture is established. The restored image $\hat{b}$ is downsampled by antialiased bicubic interpolation and refined by the generator, producing estimated radiance maps $\{\hat{b}^L\}$, where $L = 0, 1$ represent coarse and fine scales respectively. The discriminators $D = \{D^L | L = 0, 1\}$ determine whether the estimate is real or false on each scale. The adversarial objective at this stage is written as: 
\begin{subequations}
\begin{alignat}{2}
\mathcal{L}_{GAN, G}&=-\lambda_{ADV}\cdot \sum_{L}\mathbb{E}[D^{L}(\hat{b}^L)]\\
\mathcal{L}_{GAN,D}
&= \lambda_{ADV}\cdot \sum_{L}\mathbb{E}[D_2^L(b^L)]-\mathbb{E}[D^{L}(\hat{b}^L)]\nonumber\\
&-\lambda_{GP}\cdot\sum_{L} \mathbb{E}[(\lVert\nabla_{\tilde{b}^L}D^L(\tilde{b}^L)\rVert_2-1)^2]
\end{alignat}
\end{subequations}
\label{eq:ObjectiveFunc_MSGAN}

\noindent where $\tilde{b}$ is sampled uniformly along a straight line between a pair of estimated and ground truth radiance maps $\hat{b}$ and $b$. $\lambda_{ADV} = 0.1$ and $\lambda_{GP} = 1$. 

The DOE (or phase mask) learning objective consists of two terms. the laser suppresion loss $\mathcal{L}_{DOE}(LSR) = \sum I_l(\lambda)/I_{l0}(\lambda)$ encourages minimizing the laser suppresion ratio, while the background suppresion loss $\mathcal{L}_{DOE}(BSR) = \sum I_{b0}(\lambda)/I_b(\lambda)$ encourages maximizing the transmission rate of non laser-induced light, such as background scene:
\begin{equation}
\mathcal{L}_{DOE}= \mathcal{L}_{DOE}(LSR) + \mathcal{L}_{DOE}(BSR)
\label{eq:RecLoss_MSGAN}
\end{equation}

The reconstruction objective consists of two terms. The first term is given by the Charbonnier $L_1$ difference between the estimated and the ground truth, where the fine-scale ground truth image is downsampled from the coarse-scale ground truth image using an anti-aliasing bicubic method \cite{hu2006adaptive}. In this method, the high-frequency components that cause aliasing artifacts are filtered by a low-pass cubic kernel. The second term is a FFT objective is expressed as the sum of absolute difference between the Fourier transforms of the ground truth radiance map and estimated radiance map at fine- and coarse scales:
radiance pyramids:
\begin{equation}
\begin{split}
\mathcal{L}_{REC}= \sum_{L = 0,1}\sqrt{\lvert b^L-\hat{b}^L\rvert^2 + \epsilon}
+\lvert \mathcal{F}(b^L)-\mathcal{F}\big(\hat{b}^L)\rvert
\end{split}
\label{eq:RecLoss_MSGAN}
\end{equation}


\subsection{Two-Stage Training}

Suppressing laser irradiation while maintaining background light transmission presents two conflicting objectives. A common strategy in multi-objective learning is to jointly train a single network across tasks, exploiting shared representations for greater efficiency and performance than training tasks independently. While this can yield positive transfer, disparities in task difficulty and objective, data distribution, or optimization dynamics often lead to gradient interference and representation bias, reducing overall accuracy \cite{yu2020gradient}. To address this issue, we propose a two-stage learning method that first learns shared features through joint training of the neural DOE and image restoration neural net and then fine-tunes parameters specific to image restoration while freezing the neural DOE.

In the \textbf{first stage}, we jointly train the DOE (phase mask) and image restoration networks, keeping both the phase representation net and restoration net parameters learnable. We optimize the mask objective and adversarial term with gradient accumulation, which provides a more stable update direction when handling noisy gradients from wide laser strength variations:
\begin{equation}
\min_{G_{rep}, G_{res}}\Bigg(\bigg(\max_{D_1, D_{2}}\mathcal{L}_{\mathrm{GAN}}\bigg) + \mathcal{L}_{\mathrm{DOE}}\Bigg)
\label{eq:TotalObj_Solution}
\end{equation}

\noindent In the \textbf{second stage}, we fine-tune the image restoration network with the phase net frozen, updating only restoration weights via mini-batch optimization of reconstruction objective and adversarial terms. Gradients are reset each iteration to avoid inaccuracies from accumulation:

\begin{equation}
\min_{G_{rep}, G_{res}}\Bigg(\bigg(\max_{D_1, D_{2}}\mathcal{L}_{\mathrm{GAN}}\bigg) + \mathcal{L}_{\mathrm{Rec}}\Bigg)
\label{eq:TotalObj_Solution}
\end{equation}

\noindent In each stage, the generators and the discriminators are trained in an alternating manner. Denote $L_{GAN} = L_{GAN, G}$ and $L_{GAN} = L_{GAN, D}$ as the adversarial objectives for generators and discriminators respectively, the generators seek to minimize the adversarial and reconstruction objectives, while the discriminators $D_1$ and $D_2$ aim to maximize only the adversarial terms. Using only the generators at the inference stage, the radiance of the scene is restored in an end-to-end manner. 




\section{Experiment Setup}
\label{sec:experiments}


Deep learning systems rely on large-scale and high-quality data to achieve optimal results \cite{zhou2025opening, zhou2025mdk12}. A set of 100K unique color images of versatile contents and 4K resolution ($3860 \times 2160$) is collected\cite{unsplash2022} as RGB input scene radiance $b$ for training, and another 1K testing images are collected for testing. During training, coded sensor images are numerically simulated in an online manner. To match the resolution of our laboratory camera sensor ($2048 \times 2048$), each image is randomly cropped. The simulated sensor images are then downsampled to $256 \times 256$ using antialiased bicubic interpolation \cite{hu2006adaptive} to reduce the computational cost of neural network training.  Laser strengths $\alpha_l$ are randomly sampled from 100K predetermined values, which are uniformly distributed in the range of $[0,\text{2e6}]$. The incident angles of laser $\vec{n}_l=(n_{u}, n_{v})$ are normally sampled, with the "3-sigma" (three times the standard deviation) set to $0.36\cdot[f/W_s, f/H_s]$ along each axis, where $f$ is the focal length, and $W_s$ and $H_s$ are the sensor width and height respectively. The models are trained with various noise strengths, where the dark current is normally sampled with a standard deviation equal to half its mean $\mu_c = 0.002\text{e}^-$. The read noise is uniformly sampled with $\mu_r \in [350, 400]\text{e}^-$ and $\sigma_r \in [10, 11]\text{e}^-$. The Gaussian-distributed photon noise has coefficients uniformly sampled $c_1 \in [0,25\%]$ and $c_2 \in [0.9, 1.1]$. A variety of background illumination strengths $\alpha_b \in [0.3, 0.7]$ are considered. Exposure times $t$ are normally sampled with a mean of $0.1$ seconds, and the standard deviation is 0.1 times the mean. 

\subsection{Model Training}
\label{ssec: Results_dataset}
Our NeuSee system was trained on eight A100-80G GPUs. To accommodate the imaging pipeline with high-resolution hyperspectral data (2048 × 2048 × 31, batch size 8) and simultaneous training of multiple networks, we employ heterogeneous data and model parallelism for distributed training, as shown in Fig.~\ref{fig:para}. Adam optimizer with momentum $\beta_1 = 0.9, \beta_2 = 0.999$ is adopted in training. Our system is trained with learning rates of 2e-4 for weights and 4e-4 for biases. In the first stage, training runs for 10,000 iterations until the minimum LSR is reached, followed by a second stage of 20,000 iterations. Learning rates are halved after the first 2,000 iterations and further reduced by 70\% every 1,000 iterations. The biases and weights are respectively zero and orthogonally initialized.

\begin{figure}[t] 
	\begin{center} 
	\includegraphics[width=1.0\linewidth]{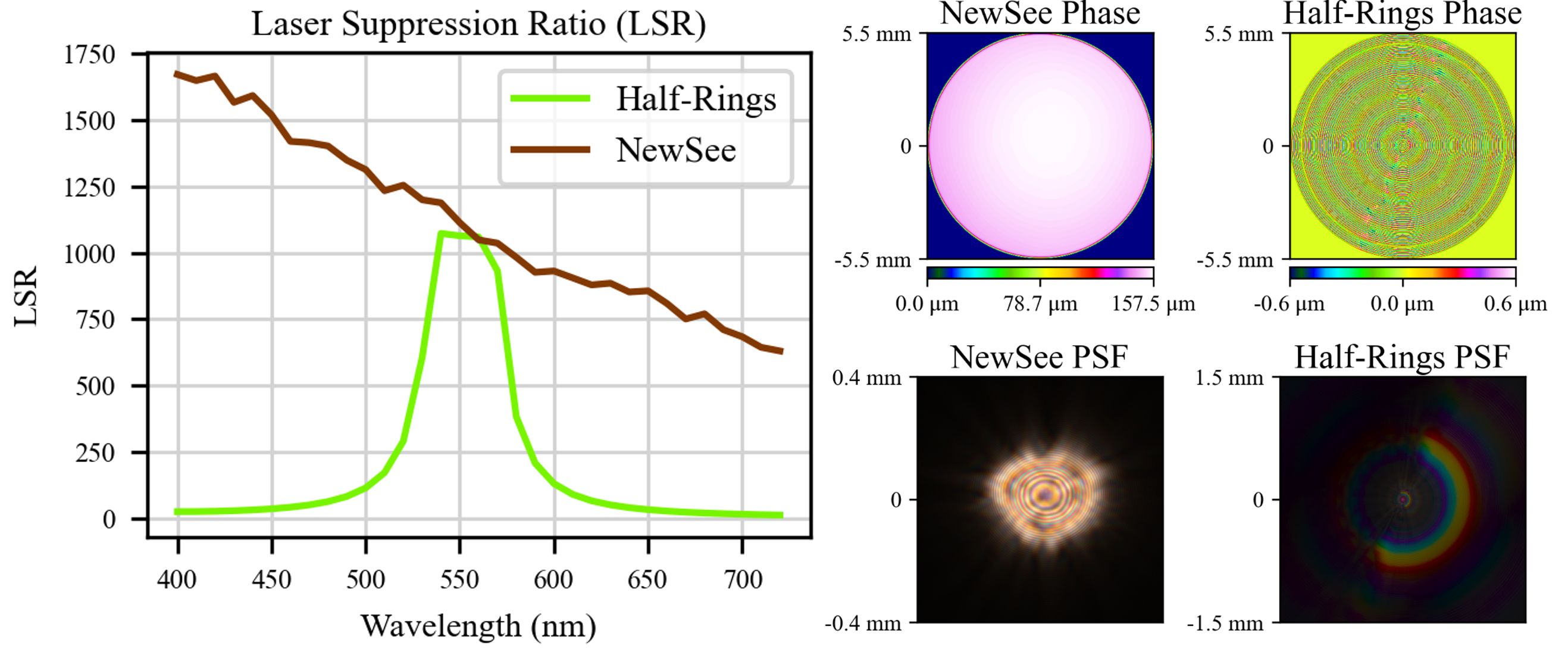} 
	\end{center} 
    \vspace{-0.5cm}
	\caption{Comparison of laser suppression ratio, DOE (or phase mask), and PSF between our jointly learned NeuSee system and the heuristically learned half-ring system \cite{wirth2020half, peng2024learning}. Our NeuSee achieves, on average, 100 times stronger suppresion of visible laser light than the half-ring.} 
    \vspace{-0.5cm}
	\label{fig:LSR}
\end{figure}

\section{Results and Discussion}
\label{subsec:main_table_results}
Fig.~\ref{fig:LSR} compares the laser suppression ratio, phase mask, and PSF of our NeuSee system with the half-ring mask \cite{wirth2020half, peng2024learning}, which was optimized using a genetic algorithm. Our NeuSee achieves a suppression over a broader laser spectral, which on average provides 5$\times$ stronger suppresion of visible laser light than the half-ring mask.

\begin{figure*}[h] 
	\begin{center} 
	\includegraphics[width=0.9\linewidth]{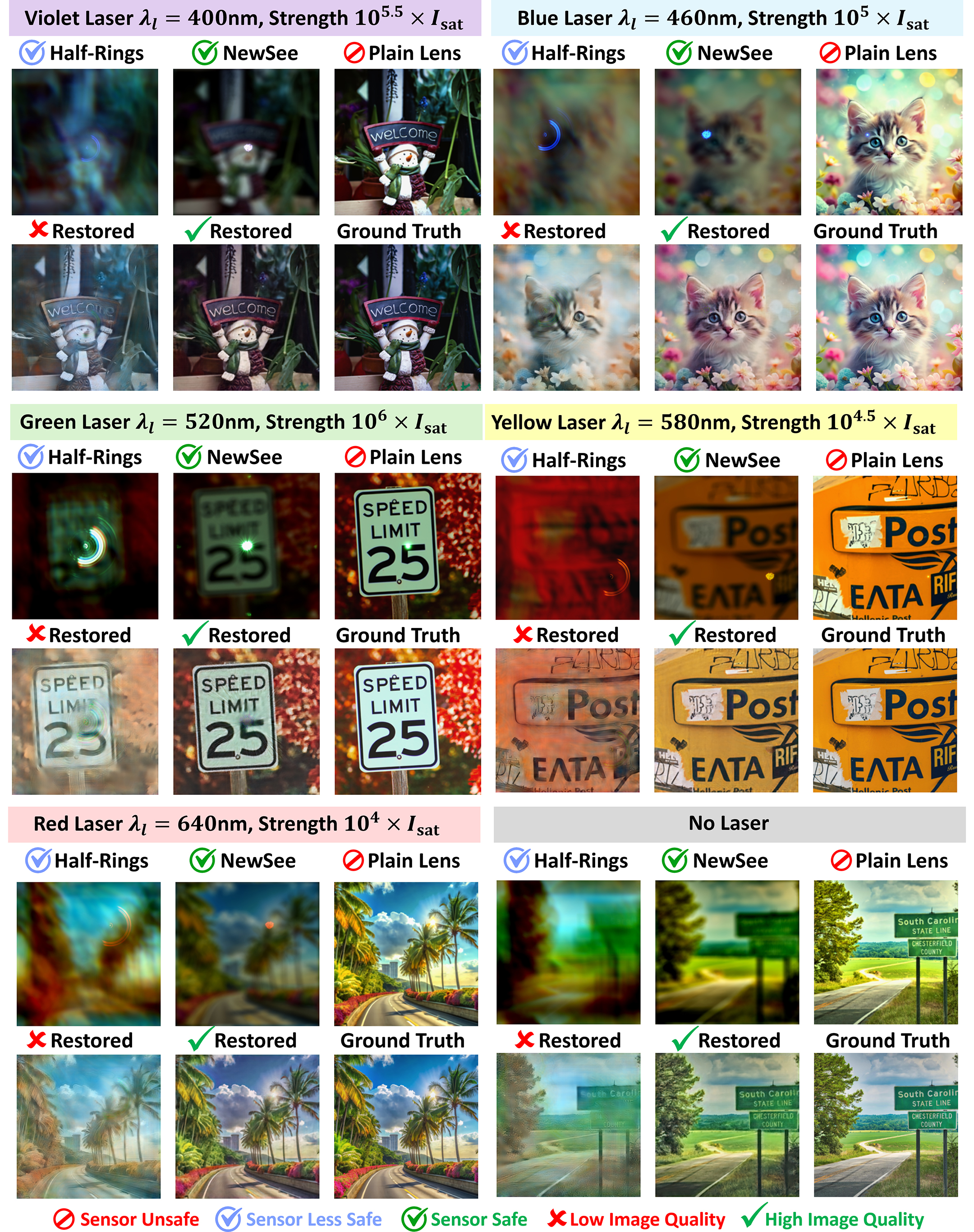} 
	\end{center} 
    \vspace{-0.5cm}
	\caption{Row 1 compares sensor image of NeuSee (learned in stage-1), the heuristically learned half-ring DOE \cite{peng2024learning}, and plain lens (lens without mask) systems. Row 2 compares image quality between NeuSee (two-stage learning) and half-ring (Mamba-GAN restored) against the corresponding ground truth images. Six lasers (10 nm full width at half maximum, 0.1\% floor noise) are presented. Our NeuSee outperforms the half-ring DOE in not only better laser suppresion and sensor protection but also  higher-fidelity image. } 
	\label{fig:resNum}
\end{figure*}

Qualitative evaluation of our NeuSee imaging is shown in Figure \ref{fig:resNum}, where the proposed NeuSee results are compared with a traditional half-ring DOE that is learned using heuristic methods \cite{peng2024learning}. Background illumination strength $\alpha_b = 0.7$, coefficients of photon noise $c_1 = 20\%$ and $c_2 = 1.0$, dark current noise $\mu_c = 0.002 \text{e}^-$, as well as the read noise $\mu_r = 390 \text{e}^-$ and $\sigma_r = 10.5 \text{e}^-$ remain the same in simulation. We showcase the presence of five lasers at voilet, blue, green, yellow, red bands. Each has a different strength, ranging from mild $\alpha_l = 1\text{e}4$ to potentially damaging laser dazzle ($\alpha_l = 1\text{e}6$) strengths. A laser free case is also demonstrated.  In both the laser-free and laser-dazzle cases, our NeuSee outperforms the heuristically learned half-ring phase mask and produces the consistently highest-fidelity reconstructions for the anti-dazzle imaging. Quantitative evaluations are performed on a set of 7K test images simulated from a thousand groundtruth scenes and seven laser strengths $\alpha_l = \{0, 10^k | k = 1, 2, ..., 6\}$. Other parameters follow the sampling scheme of the training set. Compared to the half-ring, our NeuSee improves the restored image quality by 10.1\% (L1 metric on average).

\section{Conclusion}
We present NeuSee, a computational imaging framework that jointly learns a neural diffractive optical element (DOE) and a frequency-space Vision Mamba-GAN network for image restoration. Trained end-to-end on 100K images, it protects sensors from laser flare and damage using a linear, broadband, and instantaneous optical mechanism. The DOE provides immediate protection, with system latency determined only by post-processing. In simulation, NeuSee achieves an irradiance dynamic range of $10^6$ over sensor saturation and remains robust under varying laser intensities, angles, ambient lighting, and noise, enabling potential applications in autonomous vehicles, security cameras, HDR imaging, and laser-safe headsets.


{
    \small
    \bibliographystyle{unsrtnat}
    \bibliography{main}
}

\end{document}